\documentclass{JHEP3}
\usepackage{amsmath}
\usepackage{cite}

\newcommand{\OO} {{\cal O}}
\newcommand{\LL} {{\cal L}}
\newcommand{\BB} {{\cal B}}

\newcommand{\bra}[1] {\left<#1\right|}
\newcommand{\ket}[1] {\left|#1\right>}
\newcommand{\braket}[2] {\left<#1\vphantom{#2}\right|
                         \left.\!\vphantom{#1}{#2}\right>}

\title{On the validity of the solution of string field theory}

\author{Ehud Fuchs and Michael Kroyter\\
Max Planck Insitut f\"ur Gravitationsphysik\\
Albert Einstein Institut\\
14476 Golm, Germany\\
\email{udif@aei.mpg.de, mikroyt@aei.mpg.de}
}

\abstract{
We analyze the realm of validity of the recently found tachyon solution
of cubic string field theory.
We find that the equation of motion holds in a non trivial way
when this solution is contracted with itself.
This calculation is needed to conclude the proof of Sen's first
conjecture.
We also find that the equation of motion holds when the tachyon or
gauge solutions are contracted among themselves.
}

\preprint{{\tt hep-th/0603195}\\AEI-2006-017}

\begin{document}

\section{Introduction}

Witten's action of cubic bosonic string field theory~\cite{Witten:1986cc},
\begin{equation}
\label{SFTaction}
S=\frac{1}{2}\braket{\Psi}{Q_B\Psi}+\frac{1}{3}\braket{\Psi}{\Psi\star\Psi},
\end{equation}
leads to an innocent looking equation of motion,
\begin{equation}
\label{SFTeom}
Q_B\Psi + \Psi\star\Psi = 0\,.
\end{equation}
Yet, this equation remained unsolved for twenty years,
until recently Schnabl finally found a non trivial analytic
solution~\cite{Schnabl:2005gv}.
The complexity of this equation stems from the infinite number of
string fields, its non-linearity and the difference in forms of the linear
and non-linear terms.
Many attempts have been made in order to simplify the structure of this
equation~\cite{Bordes:1991xh,Kostelecky:2000hz,Rastelli:2001rj,Gross:2001rk,Takahashi:2001pp,Kluson:2002ex,Bars:2001ag,Douglas:2002jm,Rastelli:2001hh,Boyarsky:2002yh,Bonora:2002un,Okawa:2003cm}.

Approximate numerical solutions were found earlier using
level truncation regularization~\cite{Kostelecky:1990nt}.
Level truncation solves two issues.
First, it gives a brute force tool for handling the complexity of the
star product.
Second, it offers an elegant solution to the singularities that arise
in string field theory due to the existence of an infinite tower of
degrees of freedom.

Schnabl overcame the complexity of the equation of motion
using an intelligent choice of coordinates that simplifies the star
product and a novel gauge choice, which simultaneously simplifies the
kinetic term.
The regularization of his solution came out naturally from the derivation,
and it can be written as
\begin{equation}
\label{SchnablSol}
\Psi_{\lambda} = \lim_{N\rightarrow\infty}\bigl(\lambda^N\psi_N -
    \sum_{n=0}^N\lambda^n \partial_n\psi_n\bigr)\,.
\end{equation}
The tachyon solution is $\Psi\equiv\Psi_1$,
where it should be understood that we first set $\lambda=1$
and only then take the limit $N\rightarrow\infty$.
The $\Psi_\lambda$ with $\lambda<1$ form a one parameter family of
pure gauge solutions.

In order to prove that $\Psi_\lambda$ are indeed solutions,
Schnabl showed that they satisfy the equation of motion when
contracted with a Fock space state $\bra{\phi}$, that is,
\begin{equation}
\label{FockEOM}
\braket{\phi}{Q_B\Psi_\lambda+\Psi_\lambda\star\Psi_\lambda}=0\,.
\end{equation}
Here, we follow the standard abuse of terminology,
defining a ``Fock space'' state as a state with a
finite number of excitations.
This ``Fock space'' fails to satisfy the completness requirement
of a Hilbert space.
Moreover, it is not closed under the star product. Even the
state $\ket{3}\equiv \ket{0}\star\ket{0}$ is not part of this space.
Needless to say that neither the tachyon nor the gauge solutions
are in this space.

This raises the notorious problem of defining
``the correct space of string fields''.
This entity is not easily defined due to the absence of a
positive definite norm in this space.
Nevertheless, it is clear that such a definition is crucial.
Without a proper definition of this space even the most fundamental
questions, such as the triviality (of the cohomology) of
the kinetic operator around the
new vacuum ${\cal Q}$, are not well posed. Moreover, even the
derivation of~(\ref{SFTeom}) from~(\ref{SFTaction}) requires
the definition of a space, for which a generalization of the
fundamental lemma of the calculus of variations holds.
Lacking a definition of such a space we are led to guesswork.
Due to the existence of the second term in the action~(\ref{SFTaction}),
it is natural to expect that this space forms
a star-algebra\footnote{We consider here the space with general ghost
number, without the restriction to physical states~\cite{Thorn:1986qj}.}.
Thus, it is reasonable to require that the (integer) wedge states lie
in this space~\cite{Rastelli:2000iu,Schnabl:2002gg}.
Other clues for the form of this space can come
from various anomalies. For example, it was shown by Schnabl that
the inclusion of the so called ``unbalanced wedge states'' into
the space of string fields results in
inconsistencies~\cite{Schnabl:2002gg}. Other examples of potentially
problematic anomalies include twist, midpoint and associativity
anomalies~\cite{Horowitz:1987yz,Hata:2001wa,Bars:2002bt,Erler:2003eq}.
Some other thumb rules that were considered for deciding whether a state
is legitimate or not include a well defined local coordinate
patch~\cite{Rastelli:2000iu,Rastelli:2001vb,Rastelli:2001uv,Gaiotto:2002kf,Fuchs:2004xj,Uhlmann:2004mv}
for surface
states~\cite{LeClair:1989sp,LeClair:1989sj},
eigenvalues bounded by unity for the defining matrix of a squeezed
state~\cite{Fuchs:2002zz,Fuchs:2003wu} and more~\cite{Fuchs:2005ej}.
The asymptotic behavior of level truncation coefficients is
another criterion. In fact, one of the justifications for considering
the first term of Schnabl's tachyon
solution~(\ref{SchnablSol})\footnote{This term does not contribute to the
equation of motion, since in the limit its contraction with Fock
space states vanishes.}
is to ensure a relatively fast decay of the coefficients.

We do not know how to define the space of string fields on which we should
be able to contract the solution.
Rather, we follow the common wisdom and check the contraction of the equations
of motion with the solutions themselves~\cite{Hata:2002xm,Okawa:2003cm,Okawa:2003zc}.
That is, we show that the tachyon and gauge solutions
can be added to the list of states for which the equation of motion
holds,
\begin{equation}
\label{PsiLambda}
\braket{\Psi_{\lambda_1}}
    {Q_B\Psi_{\lambda_2}+\Psi_{\lambda_2}\star\Psi_{\lambda_2}}=0\,,
\end{equation} 
for any $\lambda_1, \lambda_2 \le 1$.
This is a non trivial generalization of~(\ref{FockEOM}) since $\Psi_\lambda$
is not in the ``Fock space''.
Moreover, the tachyon solution $\lambda=1$, involves the singular state
$\psi_{N\rightarrow\infty}$. The correlation function of
this state vanishes with any ``Fock space'' state,
yet this state is clearly not zero. For example,
\begin{equation}
\lim_{N\rightarrow \infty}\braket{\psi_N}
    {\psi_N\star\psi_N}=\frac{3\sqrt{3}}{2\pi}\,.
\end{equation}
Here, we take the limit $N\rightarrow\infty$
simultaneously for all terms.
Taking, for example, $N$ to infinity in the bra while taking $2N$ to
infinity in the ket yields a different result.

The derivation of~(\ref{PsiLambda}) also completes Schnabl's proof of
Sen's first conjecture~\cite{Sen:1999mh,Sen:1999xm}. Schnabl calculated
\begin{equation}
\braket{\Psi}{Q_B\Psi}=-\frac{3}{\pi^2}\,,
\end{equation}
and {\it assumed}~(\ref{PsiLambda}) to get,
\begin{equation}
\label{SensConj}
\frac{1}{g_o^2}\left(
\frac{1}{2}\braket{\Psi}{Q_B\Psi}
+\frac{1}{3}\braket{\Psi}{Q_B\Psi+\Psi\star\Psi} \right) =
    -\frac{1}{2\pi^2g_o^2}\,.
\end{equation}
But a priori there was no justification for assuming~(\ref{PsiLambda})
without an explicit calculation.
Our calculation shows that his assumption was nevertheless correct.

The rest of the paper is organized as follows.
In section~\ref{sec:corr}
we evaluate the correlators of three $\psi_n$'s, which form
the building blocks of Schnabl's solutions $\Psi_\lambda$.
Then, we sum the relevant combinations of such correlators in
section~\ref{sec:sum} to prove our main result, eq.~(\ref{PsiLambda}).
Finally, we offer some concluding remarks in section~\ref{sec:conc}.

While this work was nearing completion the paper~\cite{Okawa:2006vm}
appeared, which overlaps a major part of this work.

\section{The cubic term}
\label{sec:corr}

First we summarize the results from~\cite{Schnabl:2005gv}
relevant to our computation.
All results are given without proof. We refer the reader
to~\cite{Schnabl:2005gv} for further details.

A standard coordinate for the CFT computations of string theory is the
upper half plane, with the half unit circle $|z|<1$ being our coordinate patch.
For SFT computation it is useful to change coordinates to a semi infinite
cylinder of radius $\pi$, namely $C_\pi$.
The conformal map to the new coordinates is $\tilde z = \tan^{-1}z$.
Now the coordinate patch is defined by the strip $-\pi/4<\Re(\tilde z)<\pi/4$.

It is natural to mode expand primary fields on the cylinder as
\begin{equation}
\tilde\OO^h(\tilde z)=\tan\circ\OO^h(z)=\sum_{n=-\infty}^{\infty}
    \frac{\tilde\OO^h_n}{\tilde z^{n+h}}\,.
\end{equation}
In this paper we only need to mode expand the energy momentum tensor
$T_{\tilde z\tilde z}$
and the $\tilde b$ ghost
\begin{align}
\LL_n &= \tan\circ L_n =\oint \frac{d\tilde z}{2\pi i}\tilde z^{n+1}
    T_{\tilde z\tilde z}(\tilde z)\,,\\
\BB_n &= \tan\circ b_n =\oint \frac{d\tilde z}{2\pi i}\tilde z^{n+1}
    \tilde b(\tilde z)\,.
\end{align}
The $\psi_n$ terms of the solution can be written as wedge states
$\ket{n+2}$ with $\tilde c$ ghost insertions and the $\BB_0+\BB_0^\dagger$
operator,
\begin{equation}
\psi_n = \frac{2}{\pi^2}U_{n+2}^\dagger U_{n+2}\left[
    (\BB_0+\BB_0^\dagger)\tilde c(-\frac{\pi n}{4})\tilde c(\frac{\pi n}{4})
    + \frac{\pi}{2}\left(\tilde c(-\frac{\pi n}{4}) + \tilde c(\frac{\pi n}{4})
    \right) \right]\ket{0},
\end{equation}
where
\begin{equation}
U_{n+2} = \left(\frac{2}{n+2}\right)^{\LL_0}\,,\qquad
U_{n+2}^\dagger U_{n+2} = \exp\left(-\frac{n}{2}(\LL_0+\LL_0^\dagger)\right).
\end{equation}
Ignoring for a moment the $(\BB_0+\BB_0^\dagger)$ operator,
the star multiplication
of the above states is straightforward, using the equation
\begin{align}
& U_r^\dagger U_r \tilde\phi(\tilde x_1)\cdots\tilde\phi(\tilde x_n)\ket{0}\star
U_s^\dagger U_s \tilde\psi(\tilde y_1)\cdots\tilde\psi(\tilde y_m)\ket{0} = \\
& \qquad U_t^\dagger U_t \tilde\phi(\tilde x_1+\frac{\pi(s-1)}{4})\cdots
    \tilde\phi(\tilde x_n+\frac{\pi(s-1)}{4})
    \tilde\psi(\tilde y_1-\frac{\pi(r-1)}{4})\cdots
    \tilde\psi(\tilde y_m-\frac{\pi(r-1)}{4})\ket{0},
\nonumber
\end{align}
where $t=r+s-1$.
For calculating the terms with $\BB_0+\BB_0^\dagger$
we use the two combinations,
\begin{align}
\BB_{-1}^L &= \frac{1}{2}\BB_{-1}+\frac{1}{\pi}\Big(\BB_0+\BB_0^\dagger\Big)\,,\\
\BB_{-1}^R &= \frac{1}{2}\BB_{-1}-\frac{1}{\pi}\Big(\BB_0+\BB_0^\dagger\Big)\,,
\end{align}
which are useful because they obey simple relations with respect to the
star product,
\begin{align}
\label{BL}
\BB_{-1}^L(\psi_1\star\psi_2) &= (\BB_{-1}^L\psi_1)\star\psi_2\,,\\
\label{BR}
\BB_{-1}^R(\psi_1\star\psi_2) &= (-1)^{\psi_1}\psi_1\star(\BB_{-1}^R\psi_2)\,.
\end{align}
Therefore, for the state on the left (right) side of the star product we write
$\BB_0+\BB_0^\dagger$ using $\BB_{-1}^L$ $(\BB_{-1}^R)$ and $\BB_{-1}$.
We can get rid of the $\BB_{-1}$ operator using its anti-commutation relation
\begin{equation}
\{\BB_{-1},\tilde c(\tilde z)\}=1\,,
\end{equation}
and the fact that it annihilates the vacuum.
The remaining terms can be calculated thanks to
the commutation relations
\begin{equation}
[\LL_0+\LL_0^\dagger,\BB_{-1}^L] =
[\LL_0+\LL_0^\dagger,\BB_{-1}^R] = 0\,.
\end{equation}
Then the relations~(\ref{BL}, \ref{BR}) can be used again to get back a state in
the form of a wedge state with $\tilde c$ ghost insertions and the
$\BB_0+\BB_0^\dagger$ operator.
A straightforward but tedious use of the above relations leads to the result,
\begin{align}
\nonumber
\psi_n\star\psi_m\star\psi_k =
\Big(\frac{2}{\pi}\Big)^3 & U_{p+2}^\dagger U_{p+2} \left[
    \frac{1}{\pi}(\BB_0+\BB_0^\dagger)
    \tilde c(\frac{\pi p}{4}) \tilde c(-\frac{\pi p}{4}) -
    \frac{1}{2}\left(\tilde c(\frac{\pi p}{4}) +
	\tilde c(-\frac{\pi p}{4})\right)\right] \\ &\cdot
    \left(\tilde c(\frac{\pi(-n+m+k+2)}{4}) - \tilde c(-\frac{\pi(-n+m+k)}{4}\right)
    \\ 
\nonumber&\cdot
    \left(\tilde c(\frac{\pi(-n-m+k-2)}{4}) - \tilde c(-\frac{\pi(-n-m+k)}{4}\right)
    \ket{0},
\end{align}
where $p=n+m+k+2$.
What is left is to calculate the correlation function
\begin{equation}
\braket{\psi_n}{\psi_m\star\psi_k} =
    \braket{I}{\psi_n\star\psi_m\star\psi_k}.
\end{equation}
For this calculation we need the relations,
\begin{equation}
U_r U_s^\dagger = U_{2+\frac{2}{r}(s-2)}^\dagger U_{2+\frac{2}{s}(r-2)}\,,
\qquad \qquad U_r U_s = U_{\frac{rs}{2}} \,,
\end{equation}
to get
\begin{equation}
\bra{I} U_{p+2}^\dagger U_{p+2} =
    \bra{0} U_{p+1}^\dagger U_{p+1} =
    \bra{0} U_{p+1} \,.
\end{equation}
Next, we use the relation
\begin{equation}
\bra{0} U_r \left(r \BB_0^\dagger +(r-2)\BB_0\right) = 0 \,,
\end{equation}
to replace the $\BB_0^\dagger$ operator with $\frac{2}{p+1}\BB_0$
and the anti-commutation relation
\begin{equation}
\{\BB_0,\tilde c(\tilde z)\} = \tilde z \,,
\end{equation}
to drag $\BB_0$ to the right until it annihilates the vacuum.
Now, we only need the correlation function of three $\tilde c$ ghost
insertions on the cylinder with radius $\frac{\pi(p+1)}{2}$,
\begin{equation}
\left< \tilde c(x) \tilde c(y) \tilde c(z) \right >
    _{C_{\frac{\pi(p+1)}{2}}} =
    \Big(\frac{p+1}{2}\Big)^3
    \sin\Big(\frac{2(x-y)}{p+1}\Big)
    \sin\Big(\frac{2(x-z)}{p+1}\Big)
    \sin\Big(\frac{2(y-z)}{p+1}\Big) \,.
\end{equation}
This gives the final result,
\begin{multline}
\braket{\psi_n}{\psi_m\star\psi_k} = \\
\frac{(p+1)^2}{\pi^3}
    \sin^2\Big(\frac{\pi}{p+1}\Big)\left(
    \sin\Big(\frac{2\pi(n+1)}{p+1}\Big)
    +\sin\Big(\frac{2\pi(m+1)}{p+1}\Big)
    +\sin\Big(\frac{2\pi(k+1)}{p+1}\Big)
    \right).
\end{multline}

\section{The equation of motion}
\label{sec:sum}

Schnabl showed that the equation of motion holds when contracted
with Fock space states.
In this section we demonstrate that it also holds when contracted
with any of the solutions.
The first term in~(\ref{SchnablSol}) is the problematic one since it is
naively zero but still contributes to the action.
We show that all combinations of the tachyon and gauge solutions
satisfy the equation of motion.
For the gauge solutions this seems redundant since they do
not have the problematic $\psi_N$ term.
Then, this calculation is just an extra check that comes for free with
the main result.

We start with the diagonal sum over constant $p=m+k+2$
\begin{equation}
\sum_{r=-p+2}^{p-2 \text{ step } 2}
    \partial_m\partial_k\braket{\psi_n}{\psi_m\star\psi_k},
\end{equation}
where $r=m-k$.
The sum can be performed after noticing that
\begin{align}
&\partial_m\partial_k\braket{\psi_n}{\psi_m\star\psi_k} =
    \partial_p^2\left( \frac{(n+p+1)^2}{\pi^3}
    \sin^2\Big(\frac{\pi}{n+p+1}\Big)
    \sin\Big(\frac{2\pi(n+1)}{n+p+1}\Big)\right) \\
&\qquad +
    ( f_{n,p}(r-2) - 2f_{n,p}(r) + f_{n,p}(r+2) 
    + f_{n,p}(-r-2) - 2f_{n,p}(-r) + f_{n,p}(r+2) )\,,\nonumber\\
&f_{n,p}(r) = \left(\partial_r^2-\partial_p^2\right)
    \left(\frac{n+p+1}{4\pi^3}
    \sin\Big(\frac{\pi(p+r)}{n+p+1}\Big)\right) \,,
\end{align}
giving
\begin{multline}
\sum_{r=-p+2}^{p-2 \text{ step } 2}
    \partial_m\partial_k\braket{\psi_n}{\psi_m\star\psi_k} = \\
    (p-1)\partial_p^2\left( \frac{(n+p+1)^2}{\pi^3}
    \sin^2\Big(\frac{\pi}{n+p+1}\Big)
    \sin\Big(\frac{2\pi(n+1)}{n+p+1}\Big) \right) \\
+  ( f_{n,p}(p) - f_{n,p}(p-2) - f_{n,p}(-p+2) + f_{n,p}(-p) )\,.
\end{multline}
Some trigonometry shows that this is equal to
the kinetic term calculated by Schnabl,
\begin{equation}
\label{match}
\partial_p\braket{\psi_n}{Q_B\psi_{p-1}} =
\sum_{r=-p+2}^{p-2 \text{ step } 2}
    \partial_m\partial_k\braket{\psi_n}{\psi_m\star\psi_k}.
\end{equation}
Therefore, when plugging $\Psi_\lambda$ into the equation of motion,
the coefficient of $\lambda^{p}$ is zero.
This result is independent of $n$. Thus,
\begin{equation}
\label{gaugeNull}
\braket{\Psi_{\lambda_1}}
    {Q_B\Psi_{\lambda_2}+\Psi_{\lambda_2}\star\Psi_{\lambda_2}}=0\,,
\end{equation}
for $\lambda_1,\lambda_2<1$.
Combining this with Schnabl's result on the vanishing of
the kinetic term for the pure gauge solution, we see that
the cubic term in the action also vanishes.

For the cubic term of the tachyon solution we need to calculate
four terms.
The first term is,
\begin{equation}
\lim_{N\rightarrow\infty}\braket{\psi_N}{\psi_N\star\psi_N} =
    \frac{3}{\pi}\sin\Big(\frac{2\pi}{3}\Big) = \frac{3\sqrt{3}}{2\pi}\,.
\end{equation}
The second term involves a sum that in the limit $N\rightarrow\infty$
can be transformed into a Riemann integral, giving exactly the same
result as the first term,
\begin{equation}
\lim_{N\rightarrow\infty}\int_0^1 dx
    \partial_x\braket{\psi_N}{\psi_N\star\psi_{x N}} =
    \frac{3\sqrt{3}}{2\pi}\,.
\end{equation}
The third term involves two sums that, again, can be transformed
into integrals,
\begin{equation}
\lim_{N\rightarrow\infty}\int_0^1 dx\int_0^1 dy
    \partial_x\partial_y\braket{\psi_N}{\psi_{x N}\star\psi_{y N}} =
    \frac{3\sqrt{3}}{2\pi}\,.
\end{equation}
The fourth term involves three sums, but these sums cannot be
transformed into an integral near the origin. Still, we calculate
the cubic integral,
\begin{equation}
\lim_{N\rightarrow\infty}\int_0^1 dx\int_0^1 dy\int_0^1 dz
    \partial_x\partial_y\partial_z
    \braket{\psi_{z N}}{\psi_{x N}\star\psi_{y N}} =
    \frac{3\sqrt{3}}{2\pi}\,,
\end{equation}
but replace the result of the integral with a sum for
the corner of the cube
$x+y+z<1$. Luckily, we already have performed this sum for the pure
gauge calculation and it is zero~(\ref{match},~\ref{gaugeNull}).
All that is left is to subtract the integral for this illegal domain,
\begin{equation}
\label{corner}
\lim_{N\rightarrow\infty}\int_0^1 dx\int_0^{1-x} dy\int_0^{1-y-x} dz
    \partial_x\partial_y\partial_z
    \braket{\psi_{z N}}{\psi_{x N}\star\psi_{y N}} = 
    \frac{3}{\pi^2}\,.
\end{equation}
Amazingly, all the other contributions cancel and we are only
left with this last contribution from the illegal domain,
\begin{equation}
\braket{\Psi}{\Psi\star\Psi} = \frac{3}{\pi^2}\,.
\end{equation}
Together with Schnabl's result for the kinetic term we
get that the equation of motion holds when contracted with
the solution itself,
\begin{equation}
\braket{\Psi}{Q_B\Psi+\Psi\star\Psi} = 0\,.
\end{equation}
This also completes the proof of Sen's first conjecture~(\ref{SensConj}).

We have also checked that the equation of motion for the tachyon
solution holds when contracted with a gauge solution and that
the equation of motion for the gauge solution holds when
contracted with the tachyon solution.

\section{Conclusions}
\label{sec:conc}

We showed that Schnabl's tachyon solution of string field theory is
valid in the sense that it solves the equation of motion even when
contracted with
itself. Lacking a definition for the space of string states, our check
gives a good indication that this state is well behaved.

It is easy to generalize our calculation to see that this result is independent
of the relative regularization of the bra and the ket states.
Instead of integrating over a cube, we now have to integrate over a box.
The result of the integral will be different, but these integrals
cancel each other.
The only non trivial contribution is from the corner of the cube,
eq.~(\ref{corner}). Now, we will take
a corner from the box, but of the same shape, giving the same contribution.
Schnabl's calculation of the kinetic term can be described in the same way
as integrating over the square and then correcting for the lower triangle.
Therefore, the kinetic and cubic terms again cancel.

To conclude, using a correct regularization for the tachyon solution is crucial
for obtaining the right state. Yet after regularization, the state we
get seems to be
well behaved under all our tests, to the extent that the regularization
we used to get this state is irrelevant.

This work bridges a gap left in~\cite{Schnabl:2005gv}.
We believe that it will prove useful for a better understanding of Schnabl's
construction and for further applications thereof.
We hope that it would also serve in the search for the space of string fields.

\section*{Acknowledgments}

We would like to thank Emil Akhmedov, Sergey Frolov, Martin Schnabl,
Ashoke Sen and Stefan Theisen for discussions.
The work of M.~K. is supported by a Minerva fellowship.
The work of E.~F. is supported by the German-Israeli foundation for
scientific research.

\newpage

\bibliography{FK}

\end{document}